\documentclass[10pt,journal]{IEEEtran}
\IEEEoverridecommandlockouts

\usepackage{cite}
\usepackage{amsmath}
\usepackage{amsthm}
\usepackage{amssymb}

\usepackage{graphicx}
\usepackage{textcomp}
\usepackage{cleveref}
\usepackage{xcolor}
\usepackage{verbatim}
\usepackage{tabularray}
\usepackage{booktabs}
\usepackage{multirow}
\usepackage{makecell}
\usepackage[font=small]{caption}    

\usepackage{algorithm}
\usepackage[noend]{algpseudocode}
\makeatletter
\def\BState{\State\hskip-\ALG@thistlm}
\makeatother
\newcommand{\E}{\mathbb{E}}

\newcommand{\R}{\mathbb{R}}

\newtheorem{definition}{Definition}

\usepackage{matlab-prettifier}
\usepackage{color}
\usepackage{mathrsfs}
\usepackage[shortlabels]{enumitem}
\makeatletter
\def\BState{\State\hskip-\ALG@thistlm}
\makeatother
\def\BibTeX{{\rm B\kern-.05em{\sc i\kern-.025em b}\kern-.08em
    T\kern-.1667em\lower.7ex\hbox{E}\kern-.125emX}}
\usepackage{subcaption}

\begin{document}

\title{{\bf\Large Bi-Level Game-Theoretic Planning of Cyber Deception for Cognitive Arbitrage}
}

\author{Ya-Ting Yang and Quanyan Zhu
\thanks{The Authors are with the Department of Electrical and Computer Engineering, New York University, Brooklyn, NY, 11201, USA; E-mail: {\tt\small \{yy4348, qz494\}@nyu.edu}.}%
}

\maketitle

\begin{abstract}
       Cognitive vulnerabilities shape human decision-making and arise primarily from two sources: (1) cognitive capabilities, which include disparities in knowledge, education, expertise, or access to information, and (2) cognitive biases, such as rational inattention, confirmation bias, and base rate neglect, which influence how individuals perceive and process information. Exploiting these vulnerabilities allows an entity with superior cognitive awareness to gain a strategic advantage, a concept referred to as cognitive arbitrage. This paper investigates how to exploit the cognitive vulnerabilities of Advanced Persistent Threat (APT) attackers and proposes cognition-aware defenses that leverage windows of superiority to counteract attacks. Specifically, the proposed bi-level cyber warfare game focuses on ``strategic-level'' design for defensive deception mechanisms, which then facilitates ``operational-level'' actions and tactical-level execution of Tactics, Techniques, and Procedures (TTPs). Game-theoretic reasoning and analysis play a significant role in the cross-echelon quantitative modeling and design of cognitive arbitrage strategies. Our numerical results demonstrate that although the defender's initial advantage diminishes over time, strategically timed and deployed deception techniques can turn a negative value for the attacker into a positive one during the planning phase, and achieve at least a 40\% improvement in total rewards during execution. This demonstrates that the defender can amplify even small initial advantages, sustain a strategic edge over the attacker, and secure long-term objectives, such as protecting critical assets throughout the attacker's lifecycle.
\end{abstract}


\section{Introduction}
In finance, arbitrage typically refers to the practice of taking advantage of price differences for the same asset across different markets to make a risk-free profit \cite{merton1973theory}. Cognitive arbitrage, which is inspired by such a financial counterpart, refers to identifying and exploiting disparities within systems that arise from human cognitive gaps \cite{bicknell2023cognitive}. Cognitive gaps may come from two key sources: (1) cognitive capabilities \cite{marsiske2013race}, which include disparities in knowledge, education, expertise, or access to relevant information, and (2) cognitive biases, such as rational inattention \cite{sims2003implications}, confirmation bias \cite{nickerson1998confirmation}, and base rate neglect \cite{benjamin2019base}, which can influence how individuals perceive and process information \cite{yang2025herd}. By identifying and exploiting these gaps, one with better cognitive capability or who does not have cognitive bias can gain a strategic advantage over those with lower capability or who have cognitive biases, effectively influencing their decision-making processes or behaviors.

\begin{figure}
    \centering
    \includegraphics[width=3.5in]{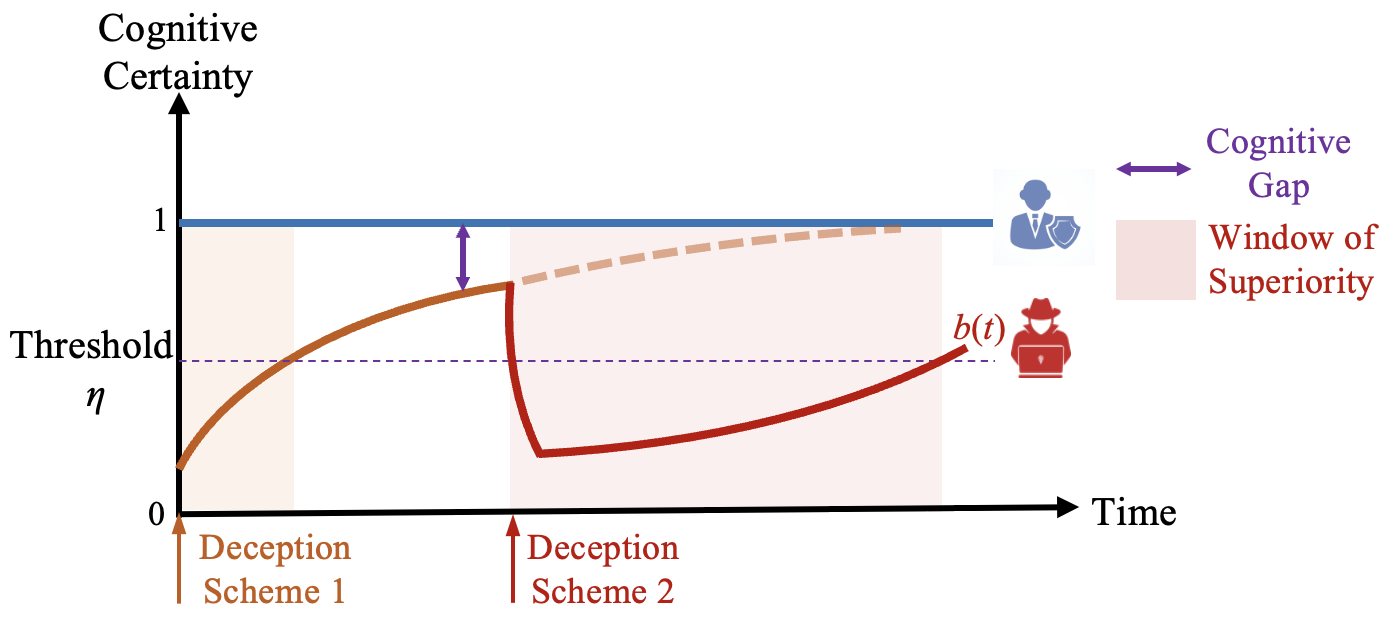}
    \caption{An illustration of cognitive arbitrage in cyber warfare. The red team's cognitive certainty in a secret (e.g., the presence of deception) often increases over time, while strategic deceptive interventions can reduce it. The double-headed arrows represent cognitive gaps, where sufficiently large gaps create superiority windows for the blue team.}
    \label{fig:fig1}
\vspace{-5mm}
\end{figure}

In adversarial settings like cybersecurity and cyber warfare \cite{kamhoua2021game}, cognitive arbitrage enables the defensive entities to exploit the malicious entities' cognitive limitations and biases for strategic advantage in the cyber battlefield \cite{tabansky2011basic}, analogous to how attackers exploit human vulnerabilities through social engineering and phishing campaigns. Since the human attacker (red team) often operates with incomplete information and is susceptible to cognitive biases, the defender (blue team) can manipulate the attacker's perception of the system secret without their awareness by deploying deception \cite{pawlick2021game,li2024symbiotic}, as illustrated in Fig. \ref{fig:fig1}. This creates cognitive gaps and superiority windows that the defender can leverage to mislead the attacker, leading them toward non-critical assets, wasting their time, and diverting their attention, while protecting critical assets. However, the attacker may gradually recognize the deception during interacting with the defender. At that point, the defender can no longer restrain the attacker unless a new trigger or deception is used. Hence, strategically designing and timing deception becomes a crucial challenge.

When designing defensive deception, several other challenges remained. One of the key issues is the dynamic nature of the problem, as the cognitive gaps may change during interactions within the dynamic system. This requires deception designs to be adaptable and capable of evolving with the system as well as the attacker's beliefs and responses, in order to ensure the defender's advantages. Another key challenge is quantifying the cognitive arbitrage capabilities of each deception technique, as there is a need for reliable metrics to assess how well (how long) each deception contributes to maintaining sufficiently large cognitive gaps between the defender and the attacker.

To address these challenges, we propose the cyber warfare game in this study, a bi-level framework, aimed at analyzing how the defender can leverage different levels of deception design \cite{li2024symbiotic} in the cyber battlefield for cognitive arbitrage. The \emph{operational level} is modeled as a one-sided information Markov game \cite{bacsar1998dynamic}, capturing the dynamic interactions and cognitive asymmetry between the defender and the attacker. Additionally, to measure the cognitive arbitrage capabilities of different deception techniques, we introduce the concept of the \emph{window of superiority} \cite{raio2023toward}, which helps quantify how long the defender maintains cognitive advantages over the attacker at the operational level, given a certain deception technique. To mitigate the chance that the attacker eventually identifies the deception, the framework includes a higher \emph{strategic level} modeled as a dynamic optimal timing problem. This level focuses on optimizing both the timing and selection of deception techniques to maintain the defender’s advantage \cite{yang2025deceive}. The proposed solution concept, cyber warfare equilibrium, for offline planning can support policy decisions, such as determining budget constraints for operations and optimal allocation of deception schemes.

Drawing on a recent cyber range human-subject research dataset that examines attackers’ cognitive vulnerabilities, we construct a numerical case study \cite{exp3} in which the embedded deception schemes have distinct levels of capability. The defender's initial cognitive advantage can diminish over time through continued interaction with the attacker. However, we show that by strategically timing the deployment of deception techniques, the defender can amplify small cognitive gaps to maintain superiority at each stage and achieve long-term goals.  The defender’s advantage diminishes as the attacker’s lifetime in the system increases. Mitigating this requires expanding the budget, deploying more sophisticated deception schemes, and strengthening intrusion detection to reduce attacker dwell time. These findings offer insights for strengthening long-term defense and informing even higher-level policy decisions. To this end, our contribution can be summarized as follows.
\begin{itemize}
    \item[(i)] We define the concept of cognitive arbitrage in cybersecurity and propose a bi-level cyber warfare game framework for holistic defensive deception that integrates both operational planning and strategic design. 
    \item[(ii)] We model attacker-defender interactions at the operational level as a one-sided information Markov game to assess the capability of cyber deception schemes, define the ``window of superiority'' to quantify deception capabilities, and formulate the strategic level as an optimal timing decision problem. 
    \item[(iii)] We establish computational foundations that support both offline planning and online execution phases of the framework. Through numerical case studies, we show how small cognitive gaps can be magnified into a substantial superiority window in strategic deception design, yielding a phenomenon analogous to the Parrondo paradox.
\end{itemize}
The rest of the paper is organized as follows. Section II reviews related work. Section III introduces the proposed bi-level cyber warfare game. Section IV presents the solution concepts and algorithms for the offline deception planning, followed by Section V, which details the execution part. Section VI provides numerical results and discusses a Parrondo-like paradox. Finally, Section VII concludes the paper.

\section{Literature Review}
\subsection{Cognitive Gaps in Decision-Making}
Cognitive biases represent thinking patterns of deviation from rational judgment, leading individuals, such as human attackers, to make irrational or sub-optimal decisions \cite{hilbert2012toward,lemay2018cognitive}. For example, \cite{simon1955behavioral} proposes replacing the optimization problem of maximizing expected utility with a simpler decision criterion known as ``satisficing'', which means that humans will stop searching for the optimal strategy once they find one that meets or exceeds their aspiration level. Besides, an experimental study \cite{cox2020stuck} reveals that people tend to self-enhance when assessing risk, believing they are less likely than others (base rate) to engage in decisions that threaten themselves. A recent study \cite{shinde2024modeling} focuses on modeling how fundamental attribution error and confirmation bias impact human beliefs as they act and obtain more information. The review \cite{berthet2022impact} gives a comprehensive introduction to other common cognitive biases, while \cite{pfleeger2012leveraging} surveys cases specifically in the context of cybersecurity. Additionally, the newly released dataset \cite{exp3} incorporates ``triggers'' specifically designed to elicit cognitive biases, enabling hypothesis-driven analyses of human (attacker) behavior and susceptibility to cognitive manipulations. The study \cite{huang2024psyborg} develops a multi-agent cybersecurity simulation environment utilizing the hidden Markov model to capture human decision-making processes under rational conditions as well as under different cognitive vulnerabilities.

\subsection{Deception Strategies in Cybersecurity}
Cyber deception defenses play a vital role against malicious attackers seeking to breach systems and compromise data \cite{zhang2021three}. The deception defense strategies and technologies are often carefully designed to misdirect cyber attackers, thus protecting valuable assets. Conventionally, defenders often construct a parallel reality with fabricated services for decoying purposes, such as Honeypots or fictitious documents in a system \cite{franco2021survey}, making the distinction between genuine and fabricated assets exceptionally challenging for the attackers. While these techniques may seem promising in theory and simulation, their capability in real-world scenarios remains questionable, particularly against human attackers \cite{aggarwal2024discovering}. Different from the controlled simulation environment, human attackers in practice tend to adapt their strategies based on cognitive biases such as confirmation biases studied in \cite{katakwar2023attackers}, making the optimal defense approach less effective than originally anticipated \cite{du2023cyber}. To address these challenges, defenders need to incorporate human cognitive vulnerabilities when developing deceptive strategies. In the context of inducing a specific cognitive state and deceiving a human attacker, a viable approach is game-theoretic reasoning and analysis \cite{zhu2019game}. One of the examples is the signaling game, which provides a valuable framework for analyzing how human attackers make decisions when faced with defensive deceptive signals \cite{pawlick2019game}.  

\section{The Bi-level Cyber Warfare Game Framework for Cognitive Arbitrage}
In this section, we begin with an overview of the different levels involved in deception design for cognitive arbitrage. We then introduce the proposed bi-level cyber warfare game framework, which focuses on the operational and strategic levels. Together, these two levels can then inform the policy decision-making during the planning phase.

\begin{figure}
    \centering
    \includegraphics[width=3.45in]{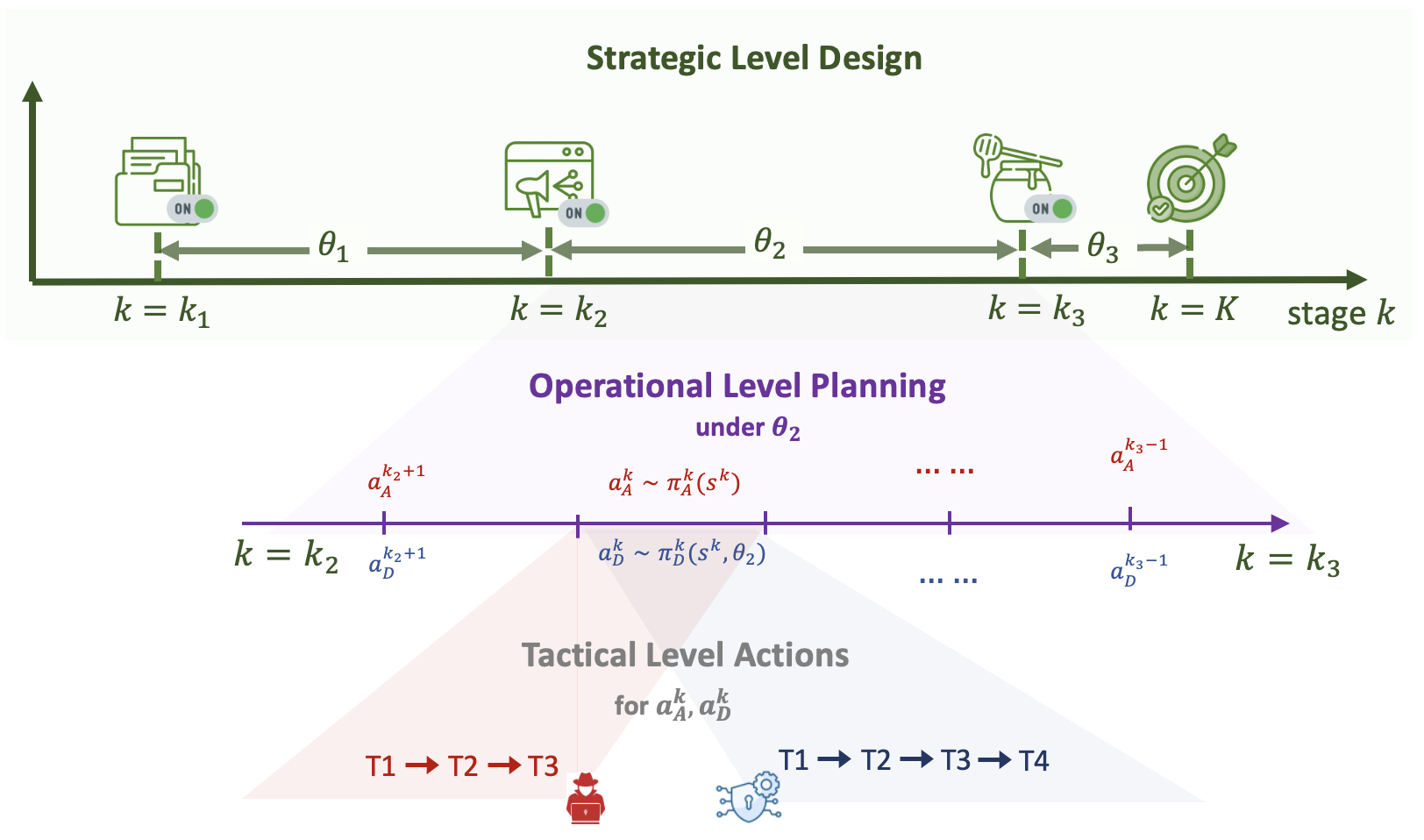}
    \caption{The strategic, operational, and tactical levels for defensive deception design. The strategic and operational levels constitute the proposed bi-level cyber warfare game.}
    \label{fig:fig2}
\vspace{-3mm}
\end{figure}

\subsection{Levels of Deception Design}
Defensive deception design can typically be categorized into three levels, as illustrated in Fig. \ref{fig:fig2}. At the detailed tactical level, specific methods such as honeypots, data camouflage, decoy files, and misinformation banners, which align with sequences of MITRE ATT\&CK \cite{strom2018mitre} TTPs can be implemented to create confusion and misdirect the attacker's efforts. The operational level focuses on engaging with the attacker in a dynamic system environment to manage responses during an ongoing interaction. At the strategic level, the focus is on determining the optimal timing and selection from a set of budgeted deception techniques to achieve long-term objectives, such as preventing access to valuable assets.

\begin{figure*}
    \centering
    \includegraphics[width=5.in]{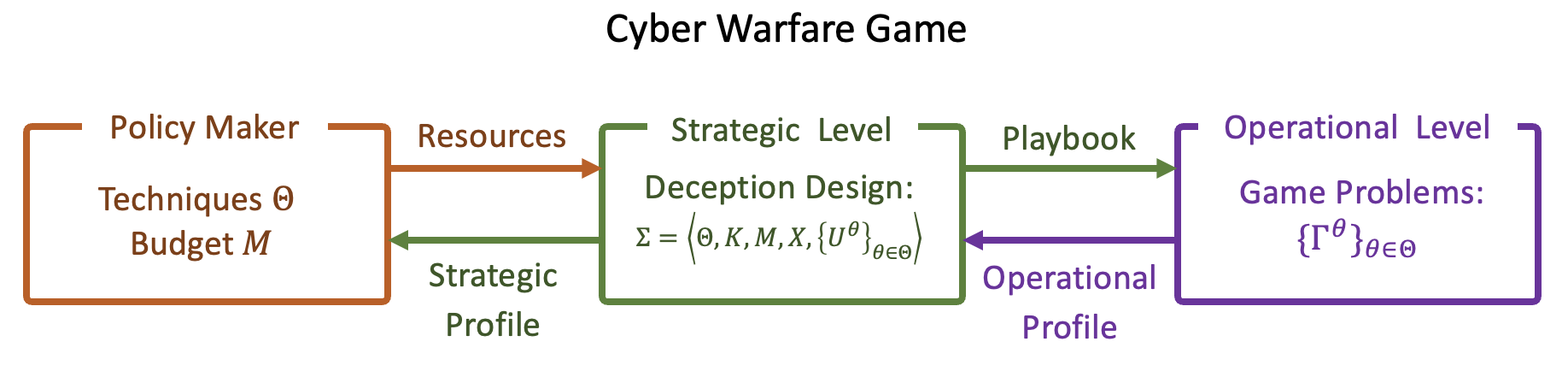}
    \caption{Conceptual architecture of the offline planning phase in the proposed bi-level cyber warfare game for deception-driven cognitive arbitrage. Policymakers define available deception techniques and budget constraints, which inform strategic-level design to optimize timing and selection of deception. The resulting playbook guides operational-level game models, whose performance feedback refines strategic and policy decisions.
}
    \label{fig:planning}
\vspace{-5mm}
\end{figure*}

Fig. \ref{fig:planning} introduces the architecture of the proposed cyber warfare game during the offline planning phase. 
The \textbf{policy maker} defines the constraints, such as the set of available deception techniques $\Theta$ and the total switching budget $M$. These resources are passed to the \textbf{strategic level}, which constitutes the cognitive arbitrage component $\Sigma$ to formulate the deception design problem as a stopping-time decision process with the objective of scheduling and selecting deception techniques over a finite horizon \(K\) to maximize long-term cognitive advantage against the adversary. The output is a deception \emph{playbook}, which is used to guide operations at the \textbf{operational level}.
At the operational level, a family of deception-aware Markov games \(\{\Gamma^\theta\}_{\theta \in \Theta}\) is used to model attacker-defender interactions under asymmetric information, incorporating evolving beliefs and responses by the attacker.
Cross-echelon decision-making profile is integrated into the architecture. The strategic designer is aware of the \emph{operational profile}: it incorporates performance evaluations of deception techniques in the operational context. Likewise, the policy maker is aware of the \emph{strategic profile}: it considers the overall expected impact of deception resource allocations over time and designs timing and selection accordingly. This backward feedback loop enables decisions to be informed by downstream performance and outcomes.
The forward arrows indicate the planning flow: how strategic-level designs are driven by policy constraints, and how operational-level actions are planned and executed in response to strategic-level guidance.

\subsection{The Operational Level}

The operational level decomposes the objective of attacking and defending selected assets into multiple stages of operations. Each game at the operational level focuses on the ongoing execution of only a specific given technique while interacting and adapting to the dynamic behavior of attackers within the system. At this level, defenders can monitor and analyze the cognitive arbitrage capability of each deception technique deployed.

One critical case is where the cognitive gap between the two players (the defender and the attacker) is due to the cognitive capabilities that lead to the attacker's unawareness (of the system secret or deception) in the end. Since unawareness represents a form of information asymmetry, we model this interactive dynamic decision-making between the defender (she) and the attacker (he) using a Markov game with one-sided information, where \textit{the defender possesses complete access to the game information, while the attacker remains unaware of the game mode representing the activated deception technique}. The game unfolds in discrete time, progressing through stages denoted by $k = 0, 1, \cdots, K$, with $K$ being the last stage of the attacker's lifetime within the system.

\begin{definition}[Operational Game (OG)]
\label{def:BG}
    The Operational Game is defined as a set of one-sided information Markov Games $\{\Gamma^{\theta}\}_{\theta \in \Theta}$ between the defender and the attacker, where $\Theta$ denotes the set of possible game modes. Each game $\Gamma^\theta$ can be represented by a tuple $\Gamma^\theta=\langle \mathcal{N}, \mathcal{S}, \theta, \{\mathcal{A}_i\}_{i \in \mathcal{N}}, T^\theta, \{r_i^\theta\}_{i \in \mathcal{N}}, \gamma, \{\mathcal{I}_i\}_{i \in \mathcal{N}} \rangle$, with components defined as follows:

    \begin{itemize}
        \item $\mathcal{N}=\{D, A\}$ is the set of players, where $P_D$ denotes the defender, and $P_A$ denotes the attacker. 
        \item $\mathcal{S}$ is the finite set of all possible states (such as location within the system) $s \in \mathcal{S}$, and the true state is directly observable for both $P_D$ and $P_A$.
    \item $\theta \in \Theta$ is the game mode (e.g., the deception technique deployed and active). The mode $\theta$ is known to $P_D$ but unknown to $P_A$ due to the latter’s limited cognitive capability.
        \item $\mathcal{A}_i$ represents the action set for player $P_i, i \in \mathcal{N}$.
        \item $T^\theta: \mathcal{S} \times \mathcal{A}_D \times \mathcal{A}_A \mapsto \Delta(\mathcal{S})$ is the state transition function conditional on current state and actions in game mode $\theta$, with $T^\theta(s'|s, a_D, a_A)$ representing the probability of transitioning to the next state $s' \in \mathcal{S}$ given the current state $s \in \mathcal{S}$, the defender $P_D$'s action $a_D \in \mathcal{A}_D$, and the attacker $P_A$'s action $a_A \in \mathcal{A}_A$.
        \item $r^\theta_i:\mathcal{S} \times \mathcal{A}_D \times \mathcal{A}_A \mapsto \mathbb{R}$ is the reward function for the immediate reward received for player $P_i$ when actions $a_D \in \mathcal{A}_D$ and $a_A \in \mathcal{A}_A$ are taken in state $s \in \mathcal{S}$ under game mode $\theta \in \Theta$. 
        \item $\gamma \in [0, 1]$ is the discount factor that determines the importance of future rewards.
        \item $\mathcal{I}_i$ is the information structure, where $\mathcal{I}_i^k$ represents the information available to Player $P_i$ at stage $k$.
    \end{itemize}
\end{definition}
The game starts in an initial state $s^0 \in \mathcal{S}$. At stage $k$ with state $s^k$, the players $P_D$ and $P_A$ select actions $a_D^k \in \mathcal{A}_D$ and $a_A^k \in \mathcal{A}_A$ based on their information structure $\mathcal{I}_D^k$ and $\mathcal{I}_A^k$, respectively. The next state $s^{k+1} \sim T^\theta(s^k, a_D^k, a_A^k)$ is then determined by the transition function associated with the current game mode $\theta$. At the same time, player $P_D$ receives $r_D^\theta(s^k, a_D^k, a_A^k)$ while player $P_A$ receives $r_A^\theta(s^k, a_D^k, a_A^k)$. 

To optimize their policy against the opponent, players can take advantage of all available information up to the point of decision-making, which is known as the behavioral policy. Hence, at stage $k$, each player's policy is a mapping from the information structure to the distribution of actions: $\pi_i^k:\mathcal{I}_i^k \mapsto \Delta(\mathcal{A}_i), i \in \mathcal{N}$, with $\pi_i = [\pi_i^0, \cdots, \pi_i^k, \cdots, \pi_i^{K}], i \in \mathcal{N}$. In this work, we specifically focus on the Markov (mixed) policy, a particular type of behavioral policy, for both players. Since the defender $P_D$ has complete information, in the Markov setting, $\mathcal{I}_D^k=\{s^k, \theta\}$, we have $\pi_D^k: \mathcal{S} \times \Theta \mapsto \Delta(\mathcal{A}_D)$. However, for the attacker $P_A$, the game mode $\theta$ is unobservable due to the unawareness of the deception technique. The attacker can only observe the true state $s^k$, so his information set is $\mathcal{I}_A^k=\{s^k\}$, then his policy is defined as $\pi_A^k: \mathcal{S} \mapsto \Delta(\mathcal{A}_A)$. Note that the true game mode $\theta$ is not directly observable to the attacker, the attacker in this case can only maintain a belief of the current game mode $b^k \in \Delta(\Theta)$, which is a probability distribution over all possible game modes, with $b^k(\theta)$ representing the attacker's belief of being in mode $\theta \in \Theta$ at stage $k$. A typical method for a rational attacker to update their belief is through the Bayes rule, as expressed in \eqref{eq:b_update}.
\begin{figure*}
\begin{equation}
        b^{k+1}(\theta|s^{k+1}) = \frac{\sum_{a_D, a_A} T^{\theta}(s^{k+1}|s^k, a_D, a_A) \pi_D^k(a_D|s^k, \theta)\pi_A^k(a_A|s^k)b^k(\theta|s^k)}{\sum_{a_D, a_A, \theta'} T^{\theta'}(s^{k+1}|s^k, a_D, a_A) \pi_D^k(a_D|s^k, \theta')\pi_A^k(a_A|s^k)b^k(\theta'|s^k)}.
    \label{eq:b_update}
\end{equation}
\noindent\rule{0.99\textwidth}{0.4pt}
\end{figure*}

Given the initial state $s^0 \in \mathcal{S}$ and the policy profile of both players $\pi=[\pi_D, \pi_A$], the player $P_D$ aims to maximize the expected cumulative discounted reward over a finite game horizon $K$:
\begin{equation}
    R_D(s^0, \pi_D, \pi_A, \theta) = \E_{\pi_D, \pi_A, T^\theta} \left[\sum_{k=0}^K \gamma^k r_D^\theta(s^k, a_D^k, a_A^k)\right].
\label{eq:e_reward_d}
\end{equation} 
As the attacker does not know the true game mode $\theta$, he needs to take the expectation of his cumulative discounted reward over a finite horizon $K$ with respect to $\theta$ at each stage $k$. That is, player $P_A$ intends to maximize:
\begin{equation}
    R_A(s^0, \pi_D, \pi_A, b^0) = \E_{\pi_D, \pi_A, T^\theta, b} \left[\sum_{k=0}^K \gamma^k r_A^\theta(s^k, a_D^k, a_A^k)\right],
\label{eq:e_reward_a}
\end{equation} where $b^0$ is the attacker's prior belief and $b=[b^0, \cdots, b^K]$ is a sequence of beliefs updated according to \eqref{eq:b_update}.
\vspace{2mm}

\noindent\textbf{Discussion on the impact of biases on belief updates:} It is worth noting that an attacker exhibiting cognitive biases may still intend to plan rationally and optimally during the planning phase, following the Bayes' rule for belief updates, as in \eqref{eq:b_update}. However, during the execution phase, cognitive biases can cause deviations from this rational behavior. While Bayes' rule assumes statistically optimal belief updates based on new evidence, cognitive biases may lead the attacker to process information in a suboptimal or inconsistent manner, resulting in belief updates that no longer adhere to Bayesian rationality during time-sensitive execution.

For example, an attacker with confirmation bias can persist in their initial belief or hypothesis, even when they encounter contradictory evidence or become trapped in a particular state. This can result in the attacker ignoring new information (to some extent) that suggests their policy is failing. On the other hand, an attacker suffering from base rate neglect may disregard the prior probability of certain events, focusing only on their subjective experience or specific evidence available to them at the moment. In the context of a cyber attack, this can mean that the attacker fails to take into account common knowledge or statistics they learned previously about the likelihood of success in a given policy, and instead relies exclusively on their limited perspective or recent experiences, which may be misleading or inaccurate. In the following analysis, we continue with the Bayes' case; the incorporation of cognitive biases into the attacker’s decision-making process during time-sensitive execution will be presented in \Cref{sec:biased_update}. 

Since the state transition functions, which define the probability of transitioning to the next state given the current state and actions, vary across different game modes, the defensive entity acts not only as a player but also as a system configurator. In this context, the defensive entity is responsible for deploying deception methods by selecting which $\theta \in \Theta$ to activate. By observing and analyzing the malicious agent's interactions across various environmental states under different game modes, the defensive entity can leverage this knowledge to effectively configure the system and optimize the defensive design for maintaining cognitive advantage in order to achieve a long-term objective.

\subsection{Window of Superiority for Cognitive Arbitrage}
Before delving into the strategic level, where the defensive entity determines when and which $\theta \in \Theta$ to activate, we begin by defining the concept of the Window of Superiority in the context of cognitive arbitrage, which measures the capability of a single usage of the deception technique $\theta$ at the operational level, as follows.
\begin{definition}[$\eta$-Window of Belief Superiority (WoBS)]
    Consider the Operational Game $\{\Gamma^{\theta}\}_{\theta \in \Theta}$ defined in Definition \ref{def:BG}. Given the defensive and malicious entities' policies $\pi_D$ and $\pi_A$, respectively, the $\eta$-Window of Belief Superiority (WoBS) is defined as an interval $[k_1, k_2]$ with $0 \leq k_1 \leq k_2 \leq K$ such that
    \begin{equation*}
        b^k(\theta) \leq \eta, \forall k \in [k_1, k_2],
    \end{equation*} where $\theta$ is the true game mode, $k_1$ is the time the superiority starts, $k_2$ is the time the superiority ends, and $\eta \geq 0$ is a threshold for the malicious entity's belief in the true game mode so that the defensive entity has superiority.
\label{def:e_WoS}    
\end{definition}
Hence, the $\eta$-Window of Belief Superiority ensures that the attacker's belief of the true game mode does not exceed a threshold $\eta$ between stages from $k_1$ to $k_2$.

\begin{definition}[$\zeta$-Window of Uncertainty Superiority (WoUS)]
    Consider the Operational Game $\{\Gamma^{\theta}\}_{\theta \in \Theta}$ defined in Definition \ref{def:BG}. Given the defensive and malicious entities' policies $\pi_D$ and $\pi_A$, respectively, the $\zeta$-Window of Uncertainty Superiority (WoUS) is defined as an interval $[k_1, k_2]$ with $0 \leq k_1 \leq k_2 \leq K$ such that
    \begin{equation*}
        H(b^k)=-\sum_{\theta \in \Theta}b^k(\theta)\log(b^k(\theta)) \geq \zeta, \forall k \in [k_1, k_2],
    \end{equation*} where $k_1$ is the time the superiority starts, $k_2$ is the time the superiority ends, and $\zeta \geq 0$ is a threshold for the malicious entity's uncertainty so that the defensive entity has superiority.
\label{def:e_AWoS} 
\end{definition}
Therefore, $\zeta$-Window of Uncertainty Superiority ensures that the attacker remains uncertain about the game mode with uncertainty greater than a threshold $\zeta$ between stages from $k_1$ to $k_2$. These definitions help the defender determine the duration of stages during which they maintain a cognitive advantage over the attacker. Moreover, since different deception techniques create varying lengths of windows, these definitions also aid in quantifying the cognitive arbitrage capability of each deception technique in terms of its ability to sustain cognitive advantage over the attacker's lifetime.

\subsection{The Strategic Level}
At the strategic level, the game is a coarse-grained high-level description. Building on the operational level, which helps evaluate the cognitive arbitrage capacity of each deception technique, the strategic level focuses on optimizing the timing and selection of techniques deployed for activation. It is worth noting that an attacker may fall for a specific deception once, but upon recognizing it, the cognitive gap is closed, and the attacker will not be deceived again unless the defender improves or employs a different deception scheme. Hence, previously activated deception techniques cannot be reused. Furthermore, deploying deception techniques, such as honeypots, may incur costs, making it impractical to use them without limitations. Given these constraints, we assume the defender has a limited budget for the number of times she can activate techniques and aims to strategically allocate a finite set of deception techniques to sustain cognitive advantage and keep the attacker from reaching the critical assets within the attacker's lifetime.

From the defensive perspective, consider the scenario where the defender can activate deception from a set of techniques, $\Theta=\{\theta_0, \theta_1, \cdots, \theta_{|\Theta|-1}\}$, leading to switching of the system configuration (i.e., resulting in changes of the transition probability) up to a budget of $M$ times within $K$ stages, with $M \leq K$ and each $\theta$ can only be used once. The objective of the defender is to maximize her utility by deciding the right time to switch and the right $\theta$ to take (the right deception technique to activate). Then, we can define the Cognitive Arbitrage Components (CAC).
\begin{definition}[Cognitive Arbitrage Components (CAC)]
\label{def:CAP}
    Given the Operational Game $\{\Gamma^{\theta}\}_{\theta \in \Theta}$ defined in Definition \ref{def:BG}, the Cognitive Arbitrage Components for the defender can be represented by a tuple $\Sigma=\langle \Theta,  K, M, \mathcal{X}, \{U^\theta\}_{\theta \in \Theta}\rangle$, where each element represents:
    \begin{itemize}
        \item $\Theta=\{\theta_0, \theta_1, \cdots, \theta_{|\Theta|-1}\}$ is the set for possible deception techniques of the defender. At stage $k$, the set of techniques already used is represented by $\bar{\Theta}^k \subseteq \Theta$.
        \item $K$ is the total stages of the problem, representing the attacker's lifetime within the system.
        \item $M$ is the budgeted number of times to switch (i.e., activate a deployed deception). At stage $k$, the remaining number of switches is $m^k$, with $m^k \in \mathcal{M}=\{0, \cdots, M\}$.
        \item $\mathcal{X}$ is the state space, each state $x^k=(s^k, b^k)$ at stage $k$ consists of the state of OG $s^k$ and the attacker's belief $b^k$ updated according to \eqref{eq:b_update}.
        \item $\eta$ and $\zeta$ are the thresholds for the Window of Belief and Uncertainty Superiority, respectively.
        \item $\{U^\theta\}_{\theta \in \Theta}: \mathcal{X} \times \mathcal{A}_D \times \mathcal{A}_A \mapsto \R$ are the payoff functions for the defender, with 
        \begin{equation}
        \begin{aligned}
            U^\theta(x^k, a^k_D, a^k_A) &= U^\theta(s^k, b^k, a^k_D, a^k_A)\\ &= r_D^\theta(s^k, a^k_D, a^k_A)\boldsymbol{1}_{b^k(\theta) \leq \eta},
        \end{aligned}
        \label{eq:payoff}
        \end{equation} where $\eta$ is a threshold for the attacker's belief of the current game mode $\theta^k$ at stage $k$. This utilizes the same concept for the $\eta$-Window of Belief Superiority and implies that the defender can only gain an advantage when the attacker remains uncertain or believes in the wrong game modes. Similarly, $U^\theta(s^k, b^k, a^k_D, a^k_A)= r_D^\theta(s^k, a^k_D, a^k_A)\boldsymbol{1}_{H(b^k) \geq \zeta}$ is the case where the concept for $\eta$-Window of Uncertainty Superiority is used.
    \end{itemize}
\end{definition}

\subsection{The Bi-Level Game}
Together, the OG in Definition \ref{def:BG} and the CAC in Definition \ref{def:CAP} constitute a cyber warfare game that captures decision-making for cyber deception design at both operational and strategic levels.
\begin{definition}[Cyber Warfare Game]
    Given the set of deception techniques $\Theta$ and the budget $M$, the cyber warfare game is made up of two parts: $\Omega_\Theta=\langle \{\Gamma^\theta\}_{\theta \in \Theta}, \Sigma \rangle$, where $\{\Gamma^\theta\}_{\theta \in \Theta}$ is the Operational Game as defined in Definition \ref{def:BG} for the operational level, and $\Sigma$ are the cognitive arbitrage components as defined in Definition \ref{def:CAP} for the strategic level of decision-making.
\label{def:CWG}
\end{definition}

This definition highlights the two-layered nature of the cyber warfare game.
The operational level models stage-by-stage attacker–defender interactions under a specific deception technique, accounting for information asymmetries, evolving beliefs, and bounded rationality. The strategic level oversees the engagement horizon to decide when and which deception techniques to activate, using operational performance evaluations within resource constraints. The two levels are interdependent: strategic choices set the operational game mode, while operational outcomes inform and refine strategic decisions, creating a closed-loop process that aligns tactical execution with long-term objectives.


\section{Analysis of Cyber Warfare Game}
This section introduces the solution concepts for the bi-level cyber warfare game. We begin by defining the $\epsilon$-Perfect Bayesian Nash Equilibrium ($\epsilon$-PBNE) at the operational level, followed by the introduction of the optimal switching time at the strategic level. Together, these concepts constitute the overall cyber warfare equilibrium, which serves to inform policy-oriented decision-making. We then present the algorithms that support the offline planning phase of the deception design.

\subsection{Equilibrium Policies at Operational Level}
At the operational level, the defender and the attacker aim to maximize their expected cumulative utilities, \eqref{eq:e_reward_d} and \eqref{eq:e_reward_a}, within the finite horizon $K$, by designing their policies, $\pi_D$ and $\pi_A$, respectively. As the utilities depend on the joint actions of both players, in this work, we consider one-sided $\epsilon$-Perfect Bayesian Nash Equilibrium ($\epsilon$-PBNE) as the solution concept.
\begin{definition}[One-sided $\epsilon$-PBNE]
    Consider the Operational Game $\{\Gamma^\theta\}_{\theta \in \Theta}$ defined in Definition \ref{def:BG}. Given $\epsilon \in \mathbb{R}_{\ge 0}$, a policy profile $\pi^*=[\pi_D^*, \pi_A^*]$ constitutes a one-sided $\epsilon$-Perfect Bayesian Nash Equilibrium ($\epsilon$-PBNE) if it satisfies:\\
    (C1) Belief Consistency: under policy profile $\pi^*=[\pi_D^*, \pi_A^*]$, player $P_A$'s belief $b^k$ at each stage $k=0, \cdots, K$ satisfies Bayes' rule, as in \eqref{eq:b_update}.\\
    (C2) Sequential Rationality: for all given game mode $\theta$ and initial state $s^{k_0} \in \mathcal{S}$ at every initial stage $k_0 \in \{0, \cdots, K\}$, each policy profile $\pi_i^{*, k_0:K}=[\pi_i^{k_0}, \cdots, \pi_i^K], i \in \mathcal{N}$ must be $\epsilon$-optimal in expectation, 
    \begin{align}
        &R_D(s^{k_0}, \pi_D^{*, k_0:K}, \pi_A^{*, k_0:K}, \theta) + \epsilon \nonumber\\
        &\qquad \geq R_D(s^{k_0}, \pi_D^{k_0:K}, \pi_A^{*, k_0:K}, \theta), \forall \pi_D^{k_0:K} \in \Pi_D^{k_0:K}\\
        &R_A(s^{k_0}, \pi_D^{*, k_0:K}, \pi_A^{*, k_0:K}, b^{k_0}) + \epsilon \nonumber\\
        &\qquad \geq R_A(s^{k_0}, \pi_D^{*, k_0:K}, \pi_A^{k_0:K}, b^{k_0}), \forall \pi_A^{k_0:K} \in \Pi_A^{k_0:K},
    \end{align} where $\Pi_i^{k_0:K}$ denotes the set of all policy profiles for player $P_i$ from stage $k_0$ to $K$.
    \label{def:PBNE}
\end{definition} Note that the case where $\epsilon=0$, the policy profile $\pi^*=[\pi_D^*, \pi_A^*]$ becomes an exact PBNE. Following previous studies \cite{huang2020dynamic,ge2023gazeta}, the one-sided $\epsilon$-PBNE can be found by dynamic programming and backward induction. Given the policy profile $\pi^*=[\pi_D^*, \pi_A^*]$, we can define the value function that estimates the future expected utilities for the defender and the attacker at stage $k$ in a given state $s^k$ under game mode $\theta$  as $v_D^k(s^k, \theta):=R_D(s^{k}, \pi_D^{*, k:K}, \pi_A^{*, k:K}, \theta)$ and $v_A^k(s^k):=R_A(s^{k}, \pi_D^{*, k:K}, \pi_A^{*, k:K}, b^{k})$. Then, at the final stage of the game $k=K$, given the state $s^K$, the value function for the defender is:

\begin{equation}
    v_D^K(s^K, \theta)=\sup_{\pi_D^K}\mathbb{E}_{\pi_D^K, \pi_A^{*, K}}[r^\theta_D(s^K, a_D, a_A)].
\end{equation} With belief $b^K$, the attacker's value function is:
\begin{equation}
    v_A^K(s^K)=\sup_{\pi_A^K}\mathbb{E}_{\pi_D^{*, K}, \pi_A^{K}, b^K}[r^\theta_A(s^K, a_D, a_A)].
\end{equation} 
Then, the following optimization program solves the optimal policy $\pi^{*, K}=[\pi_D^{*, K}, \pi_A^{*, K}]$ of the static (one-stage) game at the last stage for both players. 
\begin{equation}
    \begin{aligned}
        \max_{\pi_D, \pi_A, \nu_D, \nu_A} &\sum_{\theta \in \Theta} \alpha_\theta \bigg[\nu_D(s^K, \theta)+\mathbb{E}_{\pi_D, \pi_A}[r^\theta_D(s^K, a_D, a_A)]\bigg] \\
        & \qquad +\nu_A(s^K) + \mathbb{E}_{\pi_D, \pi_A, b}[r^\theta_A(s^K, a_D, a_A)]\\
        \text{s.t. } \mathbb{E}_{\pi_A}\Big[&r^\theta_D(s^K, a_D, a_A)\Big] \leq -\nu_D(s^K, \theta),\\ 
        & \qquad \qquad \qquad \qquad \qquad \forall a_D \in \mathcal{A}_D, \forall \theta \in \Theta, \ \ \textrm{(9a)}\\
        \quad \sum_{a_D \in \mathcal{A}_D} & \pi_D(a_D|s^K, \theta) = 1, \forall \theta \in \Theta,\\
        & \pi_D(a_D|s^K, \theta) \geq 0, \forall a_D \in \mathcal{A}_D,\forall \theta \in \Theta,\\
        \mathbb{E}_{\pi_D, b}&\Big[r^\theta_A(s^K, a_D, a_A)\Big] \leq -\nu_A(s^K), \forall a_A \in \mathcal{A}_A, \  \ \ \textrm{(9b)} \\
        \sum_{a_A \in \mathcal{A}_A} &\pi_A(a_A|s^K) = 1, \pi_A(a_A|s^K) \geq 0, \forall a_A \in \mathcal{A}_A,
    \end{aligned}
\label{prob:K}
\end{equation} where $\nu_D(s^K, \theta), \forall \theta \in \Theta$ and $\nu_A(s^K)$ are scalar decision variables; the non-decision variables $\alpha_\theta, \forall \theta \in \Theta$ can be any strictly positive and finite numbers. Following \cite{huang2020dynamic}, the solution to the optimization problem \eqref{prob:K} exists and is achieved at the equality of constraints (9a) and (9b). That is, we have $\nu_D^*(s^K, \theta)=-v_D^K(s^K, \theta)$ and $\nu_A^*(s^K)=-v_A^K(s^K)$. 

From stage $k=K-1$ to $k=0$, each player needs to optimize the sum of the immediate utility and the utility-to-go. Therefore, we can use the following recursive system equations to find the equilibrium policies $\pi_D^{*, k}, \pi_A^{*, k}$ backward.
\begin{align}
    v_D^k(s^k, \theta)=\sup_{\pi_D^k}\mathbb{E}_{\pi_D^k, \pi_A^{*, k}, T^\theta}[r^\theta_D(s^k, a_D, a_A) + v_D^{k+1}(s^{k+1}, \theta)], \nonumber\\
    v_A^k(s^k)=\sup_{\pi_A^k}\mathbb{E}_{\pi_D^{*, k}, \pi_A^{k}, b^k, T^\theta}[r^\theta_A(s^k, a_D, a_A) + v_A^{k+1}(s^{k+1})].
\end{align}
Hence, at stage $k$, given the state $s^k$, the attacker's belief $b^k$, and the value functions for the next stage $v_i^{k+1}, i \in \mathcal{N}$, the following optimization program solves the optimal policy $\pi^{*, k}=[\pi_D^{*, k}, \pi_A^{*, k}]$ at stage $k$ for both players.
\begin{equation}
    \begin{aligned}
        \max_{\pi_D, \pi_A, \nu_D, \nu_A} &\sum_{\theta \in \Theta} \alpha_\theta \Big[\nu_D(s^k, \theta)+\mathbb{E}_{\pi_D, \pi_A, T^\theta}[r^\theta_D(s^k, a_D, a_A)\\ & \qquad \qquad + v_D^{k+1}(s^{k+1}, \theta)]\Big] +\nu_A(s^k) \\
        & \quad + \mathbb{E}_{\pi_D, \pi_A, T^\theta, b}[r^\theta_A(s^k, a_D, a_A)+ v_A^{k+1}(s^{k+1})]\\
        \text{s.t. } \mathbb{E}_{\pi_A,T^\theta}\Big[&r^\theta_D(s^k, a_D, a_A) + v_D^{k+1}(s^{k+1}, \theta)\Big] \leq -\nu_D(s^k, \theta),\\
        & \qquad \qquad \forall a_D \in \mathcal{A}_D, \forall \theta \in \Theta, \ \hspace{18mm}  (11a)\\
        \quad \sum_{a_D \in \mathcal{A}_D} & \pi_D(a_D|s^k, \theta) = 1, \forall \theta \in \Theta,\\
        & \pi_D(a_D|s^k, \theta) \geq 0, \forall a_D \in \mathcal{A}_D,\forall \theta \in \Theta,\\
        \mathbb{E}_{\pi_D, b, T^\theta}&\Big[r^\theta_A(s^k, a_D, a_A)+v_A^{k+1}(s^{k+1})\Big] \leq -\nu_A(s^k), \\
        &\qquad \qquad \forall a_A \in \mathcal{A}_A, \ \hspace{31mm} (11b) \\
        \sum_{a_A \in \mathcal{A}_A} &\pi_A(a_A|s^k) = 1, \pi_A(a_A|s^k) \geq 0, \forall a_A \in \mathcal{A}_A.
    \end{aligned}
\label{prob:k}
\end{equation} Similarly, $\nu_D(s^k, \theta), \forall \theta \in \Theta$ and $\nu_A(s^k)$ are scalar decision variables; the non-decision variables $\alpha_\theta, \forall \theta \in \Theta$ can be any strictly positive and finite numbers. The solution to the optimization problem \eqref{prob:k} exists and is achieved at the equality of constraints (11a) and (11b), as in the study \cite{huang2020dynamic}. That is, $\nu_D^*(s^k, \theta)=-v_D^k(s^k, \theta)$ and $\nu_A^*(s^k)=-v_A^k(s^k)$. Based on \eqref{prob:K} and \eqref{prob:k}, the computational algorithm that provides an approximation to find the one-sided $\epsilon$-PBNE for the operational policy profile through iteratively alternating between the forward belief update according to \eqref{eq:b_update} and the backward strategy computation in order to satisfy both (C1) and (C2) in Definition \ref{def:PBNE} can then be found in reference \cite{ge2023gazeta}.

\subsection{Playbook Design at Strategic Level} \label{sec:optimal_switch}
At the strategic level, we denote $\bar{x}^k=(x^k, \theta^k, m^k, \bar{\Theta}^k)$ as the extended state space at stage $k$ with current game mode $\theta^k$, CAC state $x^k=(s^k, b^k)$ consisting of OG state $s^k$ and attacker's belief $b^k$, $m^k$ for the number of remaining switches (to activate another unused deception technique), and $\bar{\Theta}^k$ for the set of used techniques.  
Then, the optimal switching strategy can be solved using dynamic programming by considering the problem backward from the final stage $k=K$. Given any operational policy profile (OP), $\pi=[\pi_D, \pi_A]$ and belief sequence $b$ updated according to \eqref{eq:b_update} for OG, the value function for the defender at stage $K$ is simply defined by the terminal utility $U_T:\mathcal{S} \mapsto \mathbb{R}$, stating whether the attacker reaches the state leading to severe loss or containing the critical asset.
\begin{equation}
    \begin{aligned}
        &V_\pi^K(\bar{x}^K)=U_T(s^K).
    \end{aligned}
\label{eq:v_K}
\end{equation}
As for stage $k < K$, if there are no remaining times (budgets) to switch (i.e., $m=0$), the defender must stay and remain with the current deception technique. However, if there are remaining times to switch (i.e., $m>0$), the defender can either continue with the current technique or switch to another unused one, depending on which option offers a higher value for her. Therefore, we have the value function at stage $k$ given the defender and attacker's operational policy profile $\pi=[\pi_D, \pi_A]$ and belief sequence $b$ as follows:
\begin{equation}
        V_\pi^k(\bar{x}^k)=\begin{cases}
            V^k_{\pi,stay}(\bar{x}^k), &\text{ if } m=0,\\
            \max(V^k_{\pi,stay}(\bar{x}^k), V^k_{\pi,switch}(\bar{x}^k)), &\text{ if } m>0,
        \end{cases}
\label{eq:v_k}
\end{equation} with the value of staying being
\begin{equation}
    \begin{aligned}
    &V^k_{\pi,stay}(\bar{x}^k) =V^k_{\pi,stay}(s^k, b^k, \theta^k, m^k, \bar{\Theta}^k)\\& =\sum_{a_D \in \mathcal{A}_D}\sum_{a_A \in \mathcal{A}_A}\pi_D^k(a_D^k|s^k, \theta^k)\pi_A^k(a_A^k|s^k)\bigg[U^\theta(s^k, b^k, a_D^k, a_A^k)\\& \quad +\sum_{s' \in \mathcal{S}}T^{\theta^k}(s'|s^k, a_D^k, a_A^k)V_\pi^{k+1}(s', b^{k+1}, \theta^k, m^k, \bar{\Theta}^k)\bigg],
\end{aligned}  
\label{eq:v_stay}
\end{equation} where $T^{\theta^k}$ is the transition probability associated with the current mode $\theta^k$, and the attacker's belief $b^{k+1}$ is from operational level and updated according to \eqref{eq:b_update}. The value at the extended state $\bar{x}^k$ describes the expected return starting from that state and then acting according to the given operational policy $\pi$ when the defender decides to stay in the current game mode. That is, we have $\theta^{k+1}=\theta^k, m^{k+1}=m^k, \bar{\Theta}^{k+1}=\bar{\Theta}^k$ in this case. For the value of switching, 
\begin{equation}
    \begin{aligned}
    &V^k_{\pi,switch}(\bar{x}^k) =V^k_{\pi,switch}(s^k, b^k, \theta^k, m^k, \bar{\Theta}^k)\\& =\max_{\theta \in \Theta\setminus\bar{\Theta}^k} \sum_{a_D \in \mathcal{A}_D}\sum_{a_A \in \mathcal{A}_A}\pi_D^k(a_D^k|s^k, \theta)\pi_A^k(a_A^k|s^k)\bigg[ \\& \qquad \qquad \quad U^\theta(s^k, b^k, a_D^k, a_A^k)+\sum_{s' \in \mathcal{S}} T^\theta(s'|s^k, a_D^k, a_A^k) \\& \qquad \qquad \qquad V_\pi^{k+1}(s', b^{k+1}, \theta, m^k-1, \bar{\Theta}^k\cup\{\theta\})\bigg], 
\end{aligned}  
\label{eq:v_switch}
\end{equation} where the $\max$ operator is used to identify the unused deception technique that can maximize the expected return starting from the current extended state, considering the subsequent actions that follow the given operational policy $\pi$. Similarly, $b^{k+1}$ is the belief of the attacker at the operational level and updated according to \eqref{eq:b_update}.

\begin{algorithm}
  \caption{Strategic Level Design} \label{algo:strategic}
  \begin{algorithmic}[1]
    \State\textbf{Initialize} $D$ for storing purposes
    \Function{value\_function $V_\pi^k$}{$s^k, b^k, \theta^k, m^k, \bar{\Theta}^k$} \eqref{eq:v_k}
        \If{$(s^k, b^k, \theta^k, m^k, \bar{\Theta}^k)$ in $D^k$}
            \State \Return $V_\pi^k, \theta^{k+1}$ from $D^k[(s^k, b^k, \theta^k, m^k, \bar{\Theta}^k)]$
        \EndIf
        \If{$k \geq K$}
            \State \textbf{get} $v=V_\pi^K(s^k, b^k, \theta^k, m^k, \bar{\Theta}^k)$ by \eqref{eq:v_K}
            \State \Return $v, \theta^k$
        \EndIf
        \State \textbf{get} $v'=V_{\pi, stay}^k(s^k, b^k, \theta^k, m^k, \bar{\Theta}^k)$ by \eqref{eq:v_stay}
        \If{$m>0$}
            \State \textbf{get} $v''=V_{\pi, switch}^k(s^k, b^k, \theta^k, m^k, \bar{\Theta}^k)$ by \eqref{eq:v_switch}
            \State $\qquad$ with new mode $\theta'$
            \State \textbf{obtain} $v=\max(v', v'')$
            \State \Return $v, \theta^k$ if $v == v'$ else $v, \theta'$
        \Else
            \State \Return $v', \theta^k$ 
        \EndIf
    \EndFunction
    \State\textbf{------------------------ Planning Phase ------------------------}
    \State\textbf{Input} CAC $\Sigma=\langle \Theta, K, M, \mathcal{X}, \{U^\theta\}_{\theta \in \Theta}\rangle$, threshold $\eta$,\\ $\ \ \qquad$ OP $\pi=[\pi_D, \pi_A]$ with belief sequence $b$
    \State \textbf{Given} belief $b^K$, for all $s^K, \theta^K, m^K, \bar{\Theta}^K$, compute $V_\pi^K$.
    \For{$k=K-1$ to $0$}
        \State \textbf{Compute} $V_\pi^k$  for all $s^k, \theta^k, m^k, \bar{\Theta}^k$, given belief $b^k$
        \State \textbf{add} $(s^k, b^k, \theta^k, m^k, \bar{\Theta}^k), V_\pi^k$, next mode $\theta^{k+1}$ to $D^k$
    \EndFor
    \State \Return $D$
  \end{algorithmic}
\end{algorithm}
\vspace{-3mm}

\begin{definition}[Strategic-Level Optimal Switching Strategy]
\label{def:strategic-solution}
Given an operational policy profile $\pi=[\pi_D,\pi_A]$ and a belief sequence $b$ updated via~\eqref{eq:b_update}, \emph{strategic-level optimal switching strategy} is a mapping
\(
\psi^* : \bar{\mathcal{X}} \times \{0,\dots,K\} \to \Theta
\)
that selects at each stage $k$ either $\theta^{k+1}=\theta^k$ (\emph{stay}) or $\theta^{k+1} \in \Theta \setminus \bar{\Theta}^k$ (\emph{switch}) to maximize the defender's value function~\eqref{eq:v_k} over the horizon. The induced sequence $\{\theta^k\}_{k=0}^K$ together with the associated value functions $V_\pi^k$ forms the \emph{strategic playbook}.
\end{definition}

In practice, given any operational policy profile $\pi=[\pi_D, \pi_A]$ with the corresponding belief sequence $b$ updated according to~\eqref{eq:b_update}, the strategic-level strategy can be computed using Algorithm~\ref{algo:strategic}. Since the value functions in~\eqref{eq:v_k} are defined recursively, their results are stored in a playbook to avoid redundant computations during backward-induction.

\subsection{Cyber Warfare Equilibrium}

It is worth noting that the operational level and the strategic level are not independent. When fixing the game mode $\theta \in \Theta$ over the attacker's lifetime $K$, $\pi^*=[\pi_D^*, \pi_A^*]$ found by backward computation based on \eqref{prob:K} and \eqref{prob:k} with forward belief update $b=[b^0, b^1, \cdots, b^K]$ according to Bayes' rule in \eqref{eq:b_update} for belief consistency gives the one-sided $\epsilon$-PBNE policy as defined in Definition \ref{def:PBNE} for the operational level. Then, given $\pi^*=[\pi_D^*, \pi_A^*]$ and belief sequence $b$, \eqref{eq:v_k} decides whether to switch and which deception (mode) to switch at stage $k$. The playbook 
$D=\{D^k\}_{k=0, \cdots, K}$ with 
\begin{equation}\label{Playbook}
D^k=\{(s^k, b^k, \theta^k, m^k, \bar{\Theta}^k), V_\pi^k, \theta^{k+1}\}_{ s^k \in \mathcal{S}, \theta^k \in \Theta, m^k \in \mathcal{M}, \bar{\Theta}^k \subseteq \Theta}
\end{equation} for the strategic level can then be constructed using Algorithm \ref{algo:strategic}. 
Hence, a holistic solution concept can then be defined for the proposed cyber warfare game. 

\begin{definition}[Cyber Warfare Equilibrium]\label{def:CWE}
Let $\Omega_\Theta = \langle \{\Gamma^\theta\}_{\theta \in \Theta}, \Sigma \rangle$ be the cyber warfare game as defined in Definition~\ref{def:CWG}.  
A pair \(\Phi(\Omega_\Theta) = \{ (\pi^*, b), D \} \) is said to constitute a \emph{cyber warfare equilibrium} if the following conditions hold:  
\begin{itemize}
\item[(i)]  Operational-Level Condition: The profile $\pi^* = [\pi_D^*, \pi_A^*]$ and its corresponding belief sequence $b = \{b^k\}_{k=0}^K$ satisfy the one-sided $\epsilon$-Perfect Bayesian Nash Equilibrium (one-sided $\epsilon$-PBNE) criteria as given in Definition~\ref{def:PBNE}.
\item[(ii)] {Strategic-Level Condition:}  
The playbook $D = \{ D^k \}_{k=0}^K$ is given by (\ref{Playbook}), 
where $V_\pi^k$ is the value function defined in~\eqref{eq:v_k}–\eqref{eq:v_switch} and computed via the backward-induction procedure in Algorithm~\ref{algo:strategic}.
At each stage $k$, $\theta^{k+1}$ is chosen to maximize the defender’s expected return according to the optimization problem $\Sigma$ in Section~\ref{sec:optimal_switch}.
\end{itemize}
The playbook $D$ is \emph{consistent} with $(\pi^*, b)$, in the sense that for every stage $k$ it prescribes a strategic decision:
\(
\theta^{k+1} \neq \theta^k \quad \text{(switch)}, \quad \text{or} \quad \theta^{k+1} = \theta^k \quad \text{(no switch)},
\) according to the strategic-level defined in $\Sigma$, given operational-level $(\pi^*, b)$.
\end{definition}

The solution concept proposed for the cyber warfare game assists the defender in planning and preparing for warfare at the higher policy level. Specifically, the defender can evaluate whether the set of deception techniques $\Theta$ is sufficiently effective, assess the need to increase the budgeted $M$ for activating the deployed deception, and determine if improvements to the intrusion detection system are necessary to shorten the attacker's lifetime $K$, based on the initial-stage value $V_\pi^0$.

In this context, during the offline planning phase, the one-sided $\epsilon$-PBNE policy $\pi^* = [\pi_D^*, \pi_A^*]$ and its corresponding belief sequence $b$, as defined in Definition \ref{def:PBNE}, are computed first for the operational level. This equilibrium policy profile serves as a basis for simulating and analyzing the attacker-defender interactions at each operational stage. Based on this, the strategic level plans the optimal timing with corresponding mode (activated deception), constructing a playbook with an initial stage value that assists the policy maker in evaluating the effectiveness of the overarching defensive deception design.

\section{Online Execution}
Note that once cyber warfare begins, the environment can change rapidly due to its dynamic nature, making it impractical for the defender to recompute the playbook, as defined in Definition~\ref{def:CWE}, in real time. Therefore, an adaptive approach to bi-level deception design is necessary during execution.

\subsection{Learning and Adaptation}
When strategic-level decisions change (e.g., switch/activate earlier or later), the operational level needs to adapt to the change in an agile manner. If tactical sequences or operational actions fail due to uncertainties or unexpected events, there is a need to resiliently adjust strategic-level decisions. To address this, we introduce Algorithm \ref{algo:exe} for the execution phase.

\begin{algorithm}
  \caption{Execution Scheme for Cyber Warfare Game} \label{algo:exe}
  \begin{algorithmic}[1]
    \State\textbf{Input} $\Omega_\Theta$, planned $\Phi(\Omega_\Theta)=\{(\pi^*, b), D\}$, \\ $\ \ \qquad$ threshold $\epsilon, \eta \geq 0$, horizon $K'<K$
    \State \textbf{Given} stage $k$, $s^k$, $b^k$, $\theta^k$, $m^k$, $\bar{\Theta}^k$
    \State \textbf{Refine} OP with $K'$-stages ahead $\pi^{*, k:k+K^\prime}, b^{k:k+K^\prime}$
    \State \textbf{Compute} $V_{\pi^*}^k$ with $D$, decide $\theta^{k+1}$
    \If{$\theta^{k+1} \neq \theta^k$}
        \State  $m^{k+1}=m^k-1, \bar{\Theta}^{k+1}=\bar{\Theta}^{k}\cup\{\theta^{k+1}\}$
    \Else
        \State  $m^{k+1}=m^k,\bar{\Theta}^{k+1}=\bar{\Theta}^{k}$
    \EndIf
    \State \textbf{Take} one step forward: $(a_D^k, a_A^k) \sim \pi^{*, k}_D, \pi^{*, k}_A$ 
    \State \textbf{Observe} next state: $s^{k+1} \sim T^{\theta^{k+1}}$ 
    \State \textbf{Update} belief $b^{k+1}(\theta|s^{k+1}, a_D^k, a_A^k)$ according to \eqref{eq:b_update_exe}
    \State \Return $s^{k+1}, b^{k+1}, \theta^{k+1}, m^{k+1}, \bar{\Theta}^{k+1}$
  \end{algorithmic}
\end{algorithm}
\vspace{-3mm}
\begin{equation}
    \begin{aligned}
        &b^{k+1}(\theta|s^{k+1}, a_D^k, a_A^k)\\& = \frac{T^{\theta}(s^{k+1}|s^k, a_D, a_A) \pi_D^k(a_D|s^k, \theta)\pi_A^k(a_A|s^k)b^k(\theta|s^k)}{\sum_{\theta'} T^{\theta'}(s^{k+1}|s^k, a_D, a_A) \pi_D^k(a_D|s^k, \theta')\pi_A^k(a_A|s^k)b^k(\theta'|s^k)}.
    \end{aligned}
    \label{eq:b_update_exe}
\end{equation}

Note that in the execution scheme, the belief is updated ex-post, meaning it is computed after the action is executed and the state transitions, as shown in \eqref{eq:b_update_exe}. This updated belief then serves as the prior belief for the next round (stage).

\begin{figure}
    \centering    \includegraphics[width=1.8in]{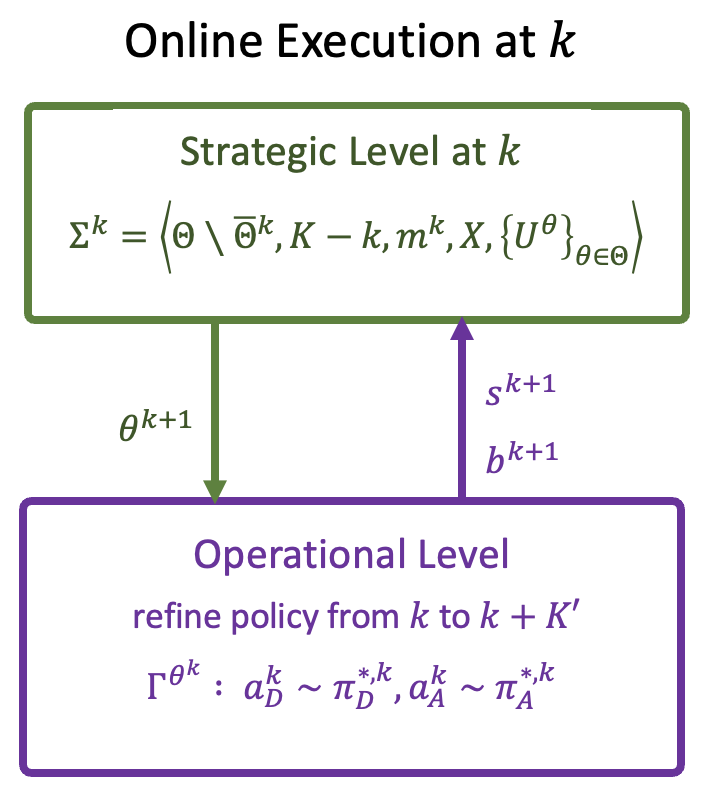}
\caption{
Bi-level interaction between the strategic and operational levels during the online execution phase of cyber deception design at stage~$k$. 
The \emph{strategic level} determines the next deception technique, while the \emph{operational level} refines and executes the corresponding policy. 
Feedback of the updated state and belief enables a closed-loop adaptation cycle between the two levels.
}
    \label{fig:execution}
\vspace{-3mm}
\end{figure}

Fig. \ref{fig:execution} then illustrates and summarizes the interdependence between the operational and strategic levels during the execution phase. At stage $k$ during the execution phase, given the horizon $K'$ for future look-ahead, the current state $s^k$, belief $b^k$, mode $\theta^k$, remaining budgeted switches $m^k$, and the set of used techniques $\bar{\Theta}^k$, the operational level refines the policy $\pi^{*, k:k+K'} = [\pi_d^{*, k:k+K'}, \pi_a^{*, k:k+K'}]$ with the consistent belief sequence $b^{k:k+K'}$. Such refinement aims to adapt the policy not only to the uncertainties and changing system environment but also to the current state $s^k$ and belief $b^k$, treating them as the new initial state and prior belief.
Then, based on the refined operational profile, the strategic level decides whether to switch modes ($\theta^{k+1} \neq \theta^k$) or maintain the current mode ($\theta^{k+1}=\theta^k$). A step is executed according to the refined operational profile, where actions $(a_D^k, a_A^k) \sim \pi^{*, k}_D, \pi^{*, k}_A$ are chosen, and the system transitions to a new state $s^{k+1}\sim T^{\theta^{k+1}}(s^k, a_D^k, a_A^k)$ according to the mode $\theta^{k+1}$ decided at the strategic level. Then, the attacker updates his belief $b^{k+1}$ according to equation \eqref{eq:b_update_exe}, and the execution proceeds to the next stage, $k+1$. Overall, the execution phase enables continuous coordination between operational decision-making and strategic mode switching, ensuring that deception design remains adaptive to both evolving system conditions as well as the attacker's responses and changing beliefs, thereby sustaining the defender's strategic advantage over the course of the engagement.

\subsection{Biased Belief Update} \label{sec:biased_update}
To incorporate cognitive biases into the attacker’s decision-making process, we define a general ex-post \textit{belief update operator} during the execution phase as follows.
\begin{definition}[Belief Update Operator]
    The belief update operator $\mathcal{B}$ during execution is defined as:
    \begin{equation}
        b^{k+1} = \mathcal{B}(b^k, s^k, a_D^k, a_A^k),
        \label{eq:b_op_exe}
    \end{equation} where $b^k$ is the belief of the attacker at stage $k$, $s^k$ is the observed state, and $a_D^k, a_A^k$ are the executed actions of the defender and attacker, respectively.
\label{def:belief_operator}
\end{definition} The specific form of $\mathcal{B}$ may vary based on the type of cognitive bias present. For instance, in the case of confirmation bias, the attacker tends to disregard new information that contradicts their prior beliefs, favoring interpretations that are consistent with their initial assumptions. This cognitive rigidity results in a slower adaptation of their belief state, even when confronted with clear and conflicting evidence. One practical modeling approach is to use a convex combination of the attacker’s previous belief and the Bayesian posterior:
\begin{equation}
b_{\text{cb}}^{k+1}(\theta) = \lambda b_{\text{cb}}^k(\theta) + (1 - \lambda) b^{k+1}(\theta)
\end{equation}
Here, the parameter $\lambda \in [0, 1]$ captures the degree of confirmation bias. A larger value of $\lambda$ reflects stronger reliance on existing beliefs, resulting in belief updates that are conservative and slow to incorporate new evidence. From a defensive perspective, this behavior can be exploited to maintain the effectiveness of deception strategies over longer durations if properly designed, as the attacker is less likely to react promptly to indicators that signal inconsistency or warning.

Another case is base rate neglect, where the attacker places excessive emphasis on the observed evidence and ignores the common prior $b^0(\theta)$ that he learned before. This can be modeled by omitting the prior term in Bayes' rule and updating beliefs based solely on the likelihood of observations:
\begin{equation}
    b_{\text{brn}}^{k+1}(\theta) \propto T^\theta(s^{k+1} \mid s^k, a_D^k, a_A^k) \, \pi_D^k(a_D^k \mid s^k, \theta) \, \pi_A^k(a_A^k \mid s^k).
\end{equation}
In this form, the attacker becomes highly reactive to recent or salient signals, which increases their vulnerability to deception that leverages misleading cues or crafted environmental noise. Although this sensitivity may accelerate the detection of real warning flags, it also creates opportunities for the defender to manipulate belief updates and induce incorrect inferences through carefully designed triggers.

These examples illustrate how to integrate cognitive modeling into adversarial reasoning frameworks. Cognitive biases such as confirmation bias and base rate neglect fundamentally alter the way attackers process information and update their beliefs. Rather than treating the attacker as a perfectly rational Bayesian agent, identifying and incorporating their cognitive biases enables a more realistic representation of the attacker's behavior. This perspective not only enhances the fidelity of behavioral simulations but also opens new avenues for strategic defense. By identifying the attacker's biases and anticipating how biased attackers are likely to misinterpret evidence or ignore relevant priors \cite{exp1,exp3}, defenders can design deception strategies at both operational and strategic levels that are not only technically sound but also psychologically effective. In particular, the ability to sustain false beliefs, misdirect attention, or cause premature action depends critically on understanding and leveraging these cognitive limitations. Ultimately, modeling attacker biases provides a foundation for shaping the informational environment in ways that impose cognitive cost, delay accurate warning recognition, and create windows of superiority for defensive advantage.

\section{Numerical Experiments}
In this section, we provide an illustrative numerical case study of an enterprise network, as illustrated in Fig. \ref{fig:net}, to evaluate the performance of the proposed bi-level deception design framework in the context of cyber warfare.

\subsection{Setups}
\begin{figure}
    \centering
    \includegraphics[width=3.3in]{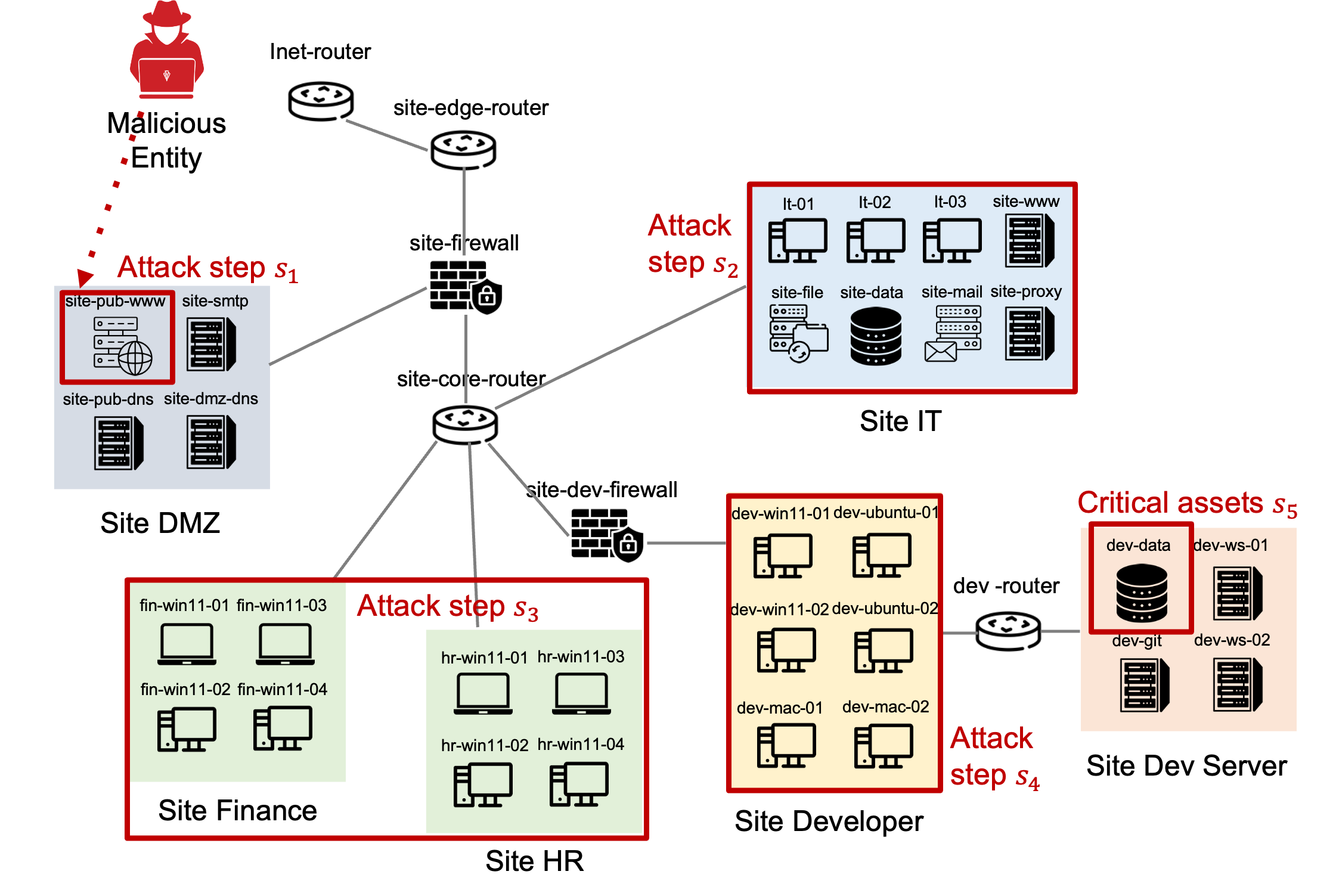}
    \caption{An example enterprise network and attack path for the case study. The path contains 5 steps (web server, site IT, site finance or HR, site developer, and site with critical asset). The attack path starts from the web server, which is open to the external network.}
    \label{fig:net}
\vspace{-3mm}
\end{figure}

Consider the scenario where a software development company or enterprise seeks to enhance the security of its network systems and protect critical assets through deception design.  
The enterprise network topology is shown in Fig.~\ref{fig:net}, with the critical assets located on the developers' server site.  
The need for interaction with the public network for external access introduces potential vulnerabilities, providing an entry point for attackers to launch advanced persistent threats (APTs).  
Typically, attackers initiate their intrusion from the external network, often targeting the web server as the first point of entry.  
The topology is consistent with the human-subject cyber range experiments described in \cite{exp1,exp3}.

An example of the attack path that a red-team agent can exploit to access the company's critical assets is also illustrated in Fig. \ref{fig:net}. At the operational level, the attacker may select actions such as exploitation, obfuscation, and privilege escalation, while the defender may implement defenses like detection, access control, and monitoring. Each of these actions involves a sequence of specific techniques and procedures at the detailed tactical level. At the strategic level, the defender may choose the timing to activate which deployed deception techniques, including banners with misinformation, fake data paths, honeypots, etc. In this case, each element of the operational game $\{\Gamma^\theta\}_{\theta \in \Theta}$ for the operational layer and the cognitive arbitrage component for the strategic layer is specified as follows.

\vspace{+2mm}
\noindent\textbf{Operational Layer:}
\begin{itemize}
    \item $s \in \mathcal{S} = \{s_1, s_2, s_3, s_4, s_5\}$, where the states correspond to the attack steps in the attack path illustrated in Fig. \ref{fig:net}.
    \item $\theta \in \Theta = \{0, 1, 2\}$, where $\theta = 0$ represents the absence of deception, $\theta = 1$ corresponds to a basic deception technique, such as a banner with misinformation, and $\theta = 2$ represents a more sophisticated deception technique, such as fake data paths.
    \item $a_D \in \mathcal{A}_D = \{0, 1, 2\}$, 
    where $a_D = 0$, $a_D = 1$, and $a_D = 2$ correspond to the defense efforts for tactics, such as intrusion detection, strong access controls, and behavioral analytics, required for $\theta = 0$, $\theta = 1$, and $\theta = 2$, respectively.
    \item $a_A \in \mathcal{A}_A = \{0, 1, 2\}$, where $a_A = 0$, $a_A = 1$, and $a_A = 2$ correspond to the attack efforts for tactics, such as standard exploitation, obfuscation, and privilege escalation, required for $\theta = 0$, $\theta = 1$, and $\theta = 2$, respectively.
\end{itemize} The detailed setups for the immediate rewards for the defender and attacker, $r^\theta_D, r^\theta_A$, and the transition function $T^\theta$ for different game modes $\theta \in \Theta$ are summarized in Table \ref{tab:IRsetup}, and \ref{tab:TPsetup}, respectively. The form of the transition probability is inspired by the experiment \cite{exp3}, in which more experienced malicious entities (experts) outperform open-division cyber attackers, and their behaviors are influenced by the presence of deception or by triggers linked to cognitive biases.

\begin{figure}
    \centering
    \includegraphics[width=3.in]{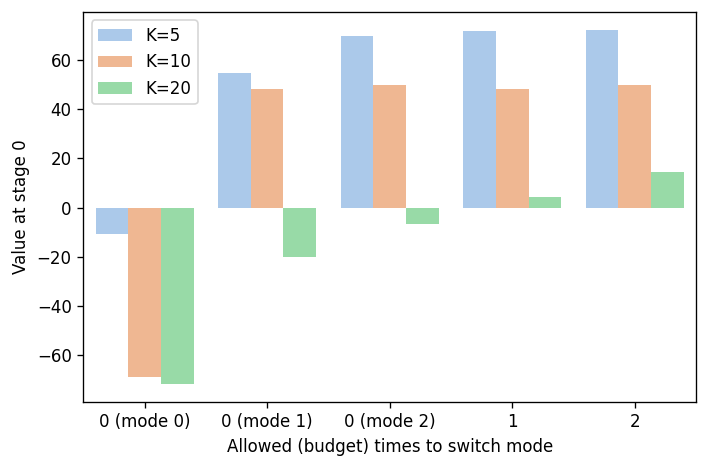}
    \caption{The results show the value at the initial stage of the defender for different budgeted switch times and varying attacker lifetimes. Three cases are analyzed for different initial game modes (modes 0, 1, and 2) when the budget is zero, and optimal switching is considered when the budget is non-zero.}
    \label{fig:v0}
\end{figure}

\vspace{+2mm}
\noindent\textbf{Strategic Layer:} For the cognitive arbitrage components $\Sigma$, the set of deception techniques $\Theta$, the state space $\mathcal{X}$, where $x^k =(s^k, b^k) \in \mathcal{X}$ follows from the operational game $\{\Gamma^\theta\}_{\theta \in \Theta}$. The total stage (attacker's lifetime) is set to $K=5, 10, 20$, and the budgeted switch is set to $M=0, 1, 2$. The payoffs $\{U^\theta\}_{\theta \in \Theta}$ for the defender follow \eqref{eq:payoff}, while the terminal rewards $U_T$ are summarized in Table \ref{tab:TRsetup}. The results of the value at the initial stage under the optimal switching strategy, considering different budgeted times for switching and varying attacker lifetimes, are shown in Fig. \ref{fig:v0}.

\begin{table}[htbp]
\caption{Immediate Rewards}
\begin{center}
\begin{tabular}{cccc}
\toprule
$a_D^k$ & $a_A^k$ & condition & $r_D^\theta(s^k, a_D^k, a_A^k)$ \\
\midrule
$\theta^k$ & \thead[l]{not $\theta^k$ \\ $\theta^k$}  & & \thead[c]{$10.0$ \\ $5.0$} \\
\midrule
not $\theta^k $ & \thead[l]{$\theta^k$ \\ not $\theta^k$ \\ }  & \thead[l]{ \\ $a_D^k > a_A^k$ \\ $a_D^k \leq a_A^k$} & \thead[c]{$0.0$ \\ $1.0$ \\ $0.0$} \\
\bottomrule \\[-0.3em]
\multicolumn{4}{c}{Here, we consider $r_A^\theta(s^k, a_D^k, a_A^k)=-r_D^\theta(s^k, a_D^k, a_A^k)$.}
\end{tabular}
\end{center}
\label{tab:IRsetup}
\end{table}

\begin{table}[htbp]
\caption{Transition Probability}
\begin{center}
\begin{tabular}{cccccc}
\toprule
$\theta^k$ & $s^k$ & $a_D^k$ & $a_A^k$ & condition & $T^\theta(s^{k+1}| s^k, a_D^k, a_A^k)$ \\
\midrule
$\theta_j$ & $s_l$ & not $\theta_j$ & $\theta_j$ &  & $\kappa-\theta_j/10 - (l/\beta)$ \\
\midrule
$\theta_j$ & $s_l$ & $\theta_j$ & $\theta_j$ &  & $\kappa-\delta-\theta_j/10 - (l/\beta)$ \\
\midrule
$\theta_j$ & $s_l$ & not $\theta_j$ & not $\theta_j$ & \thead[c]{ \\ $a_D^k > a_A^k$ \\ $a_D^k \leq a_A^k$} & \thead[c]{ \\ $\kappa-\delta$ \\ $1.0-(\kappa-\delta)$}  \\
\bottomrule \\[-0.3em]
\multicolumn{6}{c}{$\kappa$ is the attacker's ability, $\delta$ is the defender's ability, and $\beta$ denotes the}\\ 
\multicolumn{6}{c}{impact of the current state. Here, we choose $\kappa=0.8, \delta=0.5, \beta=\infty$.}
\end{tabular}
\end{center}
\label{tab:TPsetup}
\vspace{-3mm}
\end{table}

\begin{table}[htbp]
\caption{Terminal Rewards}
\begin{center}
\begin{tabular}{cccccc}
\toprule
$s^K$ & $s_1$ & $s_2$ & $s_3$ & $s_4$ & $s_5$ \\
\midrule
$U_T(s^K)$ & 100 & 50 & 10 & 0 & -100 \\
\bottomrule \\[-0.3em]
\end{tabular}
\end{center}
\label{tab:TRsetup}
\vspace{-3mm}
\end{table}

From Fig. \ref{fig:v0}, the scenario with no allowed switches and absent deception (case $0$ (mode $0$)) shows that the defender's initial cognitive advantage diminishes over time during interactions with the attacker (that is, the attacker finds out that there is no deception), as indicated by the defender's value being the lowest and negative in this case. In the case of no allowed switches with basic deception at the beginning (case $0$ (mode $1$)), the defender can expect to win when the attacker's lifetime is short. However, when the lifetime is long enough ($K=20$), the defender can still lose. The case of no-allowable switches with sophisticated deception (case $0$ (mode $2$)) exhibits a similar trend.

\begin{figure*}
    \centering
    \includegraphics[width=6.5in]{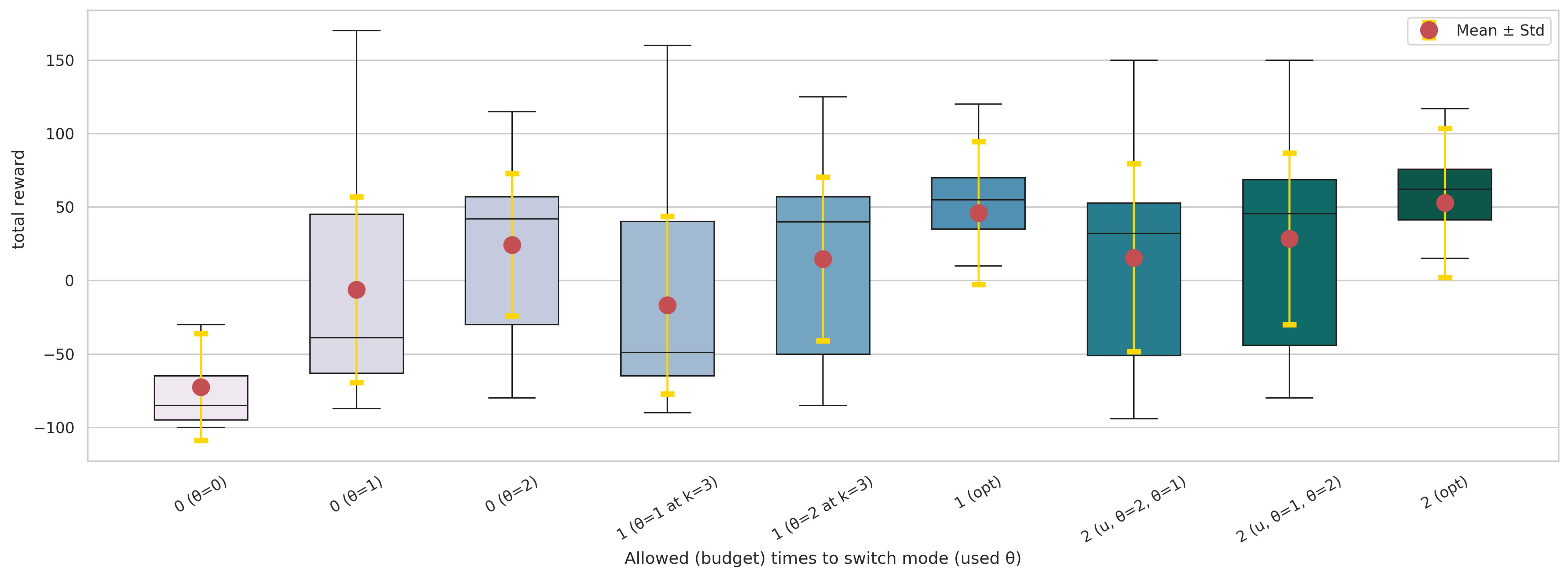}
    \caption{The results for the total rewards (sum of the terminal reward and the immediate reward at each stage) of the defender under different switching strategies. Here, the attacker's lifetime $K=10$, and each scenario is evaluated over $200$ experiments. The boxplots depict the distribution of defender rewards, with red dots representing the mean and yellow bars indicating the standard deviation across the experiments.}
    \label{fig:arbitrage_}
\end{figure*}

Then, we can observe that with a fixed attacker's lifetime, the defender's initial value increases as the budgeted number of switches increases. This indicates that strategically selecting the timing and activation of deception techniques improves the defender's performance. Additionally, we observe that the defender's value at the initial stage decreases as the attacker's lifetime increases in all five cases. This is because a longer attacker's lifetime increases the likelihood of reaching the final state, where critical assets are located. To address this, the policy maker can consider several potential solutions, including expanding the set of deception techniques, increasing the switching budget, and improving intrusion detection mechanisms to reduce the effective lifetime of the attacker.

\subsection{Comparison between different switching strategies}

To illustrate the performance of the optimal switching strategy, we consider the following baselines for comparison purposes. The results for the total rewards (sum of the terminal reward and the immediate reward at each stage) under different scenarios are shown in Fig. \ref{fig:arbitrage_}.
\begin{itemize}
    \item \textbf{No switch:} In this case, no switches are allowed; the deception technique remains fixed from the initial stage to the terminal stage. For example, in Fig. \ref{fig:arbitrage_}, case ``0 ($\theta=0$)'' denotes a scenario where the budgeted switch times are zero ($M=0$) and the deception is absent (game mode $\theta=0$). Similarly, cases ``0 ($\theta=1$)'' and ``0 ($\theta=2$)'' represent cases where no switches are allowed ($M=0$), but basic deception (game mode $\theta=1$) and sophisticated deception (game mode $\theta=2$) are employed, respectively.
    \item \textbf{Fixed timing}: In this case, the timing for activating the deployed deception (i.e., switch mode) is fixed. For instance, in Fig. \ref{fig:arbitrage_}, case ``1 ($\theta=1$ at $k=3$)'' means that the budgeted switch times is one ($M=1$) and that basic deception is activated at stage $3$ (game mode being switched to $\theta=1$ at stage $3$).
    \item \textbf{Uniform interval}: In this case, the time intervals between each deception activation are uniform. For example, in Fig. \ref{fig:arbitrage_}, case ``2 ($\theta=2, \theta=1$)'' indicates that the budgeted switch times is two ($M=2$), with sophisticated deception activated first, followed by basic deception (i.e., the game mode switches to $\theta=2$ and then $\theta=1$).
    \item \textbf{Proposed strategy}: In this case, the strategy proposed in Section \ref{sec:optimal_switch} is applied. The cases ``1 (opt)'' and ``2 (opt)'' in Fig. \ref{fig:arbitrage_} indicate that the proposed strategy is used when the budgeted switch times are one ($M=1$) and two ($M=2$), respectively.
\end{itemize}

From Fig. \ref{fig:arbitrage_}, we observe that when no switch is allowed, the deployment of sophisticated deception produces the best performance compared to basic or no deception. In the case where one switch is allowed (with the initial game mode $\theta=0$ at stage $k=0$), the proposed strategy achieves the best performance by recommending a switch to sophisticated deception at stage $1$ or $2$. The two fixed-time cases can be seen as delayed switches, with ``1 ($\theta=1$ at stage $k=3$)'' performing even worse due to selecting the wrong mode compared to the recommendation of the proposed strategy. When two switches are allowed, the proposed strategy still outperforms the uniform-interval baseline. This may be because simply applying uniform intervals between switches can lead to unnecessary early activation or delayed responses. As a result, the defender's advantage gained from the attacker's incorrect belief or high uncertainty may diminish over time, allowing the attacker to assess the situation (deception, game mode) more correctly and adjust their effort (action) accordingly. Consequently, the attacker avoids being trapped in a specific state and may ultimately reach the critical asset.

\subsection{Discussion on the Parrondo-like Paradox}
From Fig. \ref{fig:v0} in the planning phase, we observe that when the attacker's lifetime is long enough (e.g., the case where $K=20$), having $0$ budgeted switches results in a negative defender value at the initial stage, regardless of whether the initial game mode $\theta$ is $0, 1$, or $2$. This indicates that in these three game scenarios, the defender is expected to lose. However, when the budgeted switches increase to $1$, the defender's value becomes positive, and it increases further when the budgeted switches increase to $2$. This demonstrates that switching between game modes or strategically combining losing games by selecting the switching time and game mode, can lead to a winning strategy for the defender.

Similarly, in Fig. \ref{fig:arbitrage_} for the execution phase, we see that when the budgeted switch time is 0, the defender always loses (i.e., the total reward is negative) when the whole game is played in mode $\theta=0$. The defender loses more than half the time when the game mode is $\theta=1$ and loses more than a quarter of the time when the game mode is $\theta=2$. However, with optimal switching and when the budgeted switch time is $2$, the defender always wins. This shows that although the defender faces a probability of losing in each game, a winning outcome can be achieved by strategically combining the games. 

The paradoxical observations in both the planning and execution phases suggest that cognitive arbitrage or strategic deception design that involves switching between different techniques (game modes) can be more effective than relying on a single scheme. It emphasizes the importance of flexibility and adaptation in the defender's approach, allowing them to stay ahead of the attacker by exploiting their unawareness of cognitive vulnerabilities. Such an observed pattern in which losing games can be combined to form a winning strategy is the essence of the well-known Parrondo paradox, which states that pairs of games, each with a higher probability of losing than winning, can lead to winning strategies when played alternately.

\subsection{Comparison between different biases}

\begin{figure}
    \centering
    \includegraphics[width=2.8in]{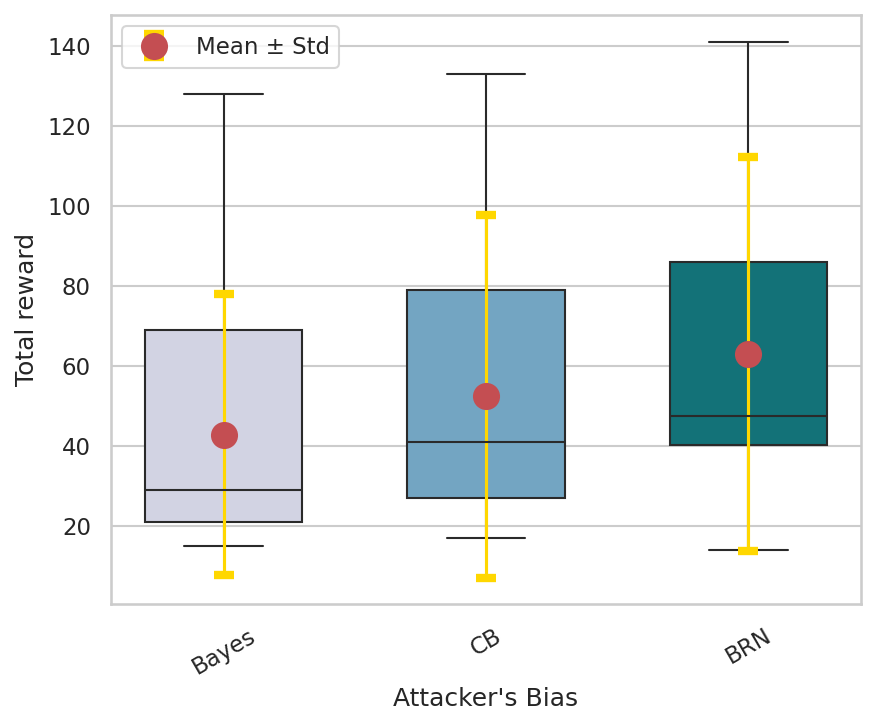}
    \caption{The results for the total rewards (sum of the terminal reward and the immediate reward at each stage) of the defender under different cognitive biases of the attacker. Here, $\lambda$ for confirmation bias is $0.5$, attacker's lifetime $K = 10$, budgeted switch times $M=2$, and each scenario is evaluated over $200$ experiments.}
    \label{fig:exp_bias}
\vspace{-3mm}
\end{figure}

Building on the discussion in Section \ref{sec:biased_update}, we consider scenarios in which the attacker exhibits confirmation bias (CB) and base rate neglect (BRN) during the execution phase, with the defender being aware of these biases and trying to strategically exploit them. The corresponding results are presented in Fig. \ref{fig:exp_bias}.

Fig. \ref{fig:exp_bias} shows that the defender gains the highest total reward when facing a BRN attacker, followed by the CB attacker, with the lowest reward observed against the Bayesian (fully rational) attacker. This suggests that biased attackers are more vulnerable to well-designed deception strategies. In the CB case, the attacker is slower to update their beliefs, allowing the defender to prolong the cognitive arbitrage capability of each deception technique. In the BRN case, the attacker heavily disregards prior information and overreacts to recent observations, making them susceptible to manipulation through saliency-driven or misleading signals.

These findings demonstrate that cognitive asymmetries can be strategically exploited by the defender to gain and sustain her advantage. More broadly, they reinforce the value of understanding, identifying and tailoring deception strategies to the attacker's cognitive vulnerabilities, highlighting how behavioral modeling can translate into improved outcomes in the context of adversarial interactions, such as cyber warfare.

\section{Conclusion}
In this work, we have proposed a bi-level cyber warfare game for defensive deception design utilizing the concept of cognitive arbitrage. The core idea is that the defender can gain a cognitive advantage through the deployment of cyber deception. However, such an advantage may diminish during the interaction with the attacker as the attacker adapts. Therefore, it is essential to dynamically determine both the optimal timing and the selection of deception to sustain the defender’s edge. At the operational level, a one-sided incomplete information Markov game has been used to model the interaction between the defender and the attacker under given deception. With the definition of cognitive (belief and uncertainty) windows of superiority, we are able to quantify the capability of each deception in terms of maintaining the defender's cognitive advantage. Then, at the strategic level, an optimal timing decision process has been formulated to optimize the deployment (or activation) timing and the selection of deception techniques to maintain the defender's superiority while preventing the attacker from reaching critical assets within the system. The results in the numerical case study have demonstrated that although the defender's initial advantage diminishes over time, strategically timed and deployed deception techniques can turn a negative value for the attacker into a positive one during the planning phase and achieve at least a 40\% improvement in total rewards during execution. This shows that the defender can amplify even small initial advantages, sustain a strategic edge over the attacker, and eventually protect critical assets throughout the attacker's lifecycle.

A possible direction for future work is to explore other types and models of the malicious entity's cognitive biases. For instance, an attacker exhibiting rational inattention or sunk cost fallacy may also deviate from Bayes' belief updating, leading to different decision-making patterns. Incorporating other biases into the framework could lead to a better understanding of how defensive deception interacts with biased attackers and provide deeper insight into the capability of different deception strategies. Another potential future direction is to consider scenarios in which the defensive entity also operates under incomplete information, specifically when the defender does not know which cognitive biases the attacker exhibits. This setting reflects a more realistic scenario in cyber operations, where the defender must infer or adapt to the adversary's cognitive state over time.

\bibliography{reference}
\bibliographystyle{IEEEtran}

\end{document}